arXiv Article

# Re-refinement of 4xan: hen egg white lysozyme with carboplatin in sodium bromide solution including details of solute, solvent, ion and split occupancy amino acid electron density evidence

Simon W. M. Tanley[1], Loes M. J. Kroon-Batenburg[2], Antoine M. M. Schreurs[2], and John R. Helliwell[1]

1.School of Chemistry, Faculty of Engineering and Physical Sciences, University of Manchester, Brunswick Street, Manchester M13 9PL, England.

2. Crystal and Structural Chemistry, Bijvoet Center for Biomolecular Research, Faculty of Science, Utrecht University, Padualaan 8, 3584 CH Utrecht, The Netherlands.

john.helliwell@manchester.ac.uk

**Abstract**

A re-refinement of 4xan, hen egg white lysozyme (HEWL) with carboplatin crystallised in NaBr solution, has been made (Tanley et al 2016). This follows our Response article (Tanley et al 2015) to the Critique article of Shabalin et al 2015, suggesting the need for corrections to some solute molecule interpretations of electron density in 4xan and removal of an organic moiety as a ligand to the platinum ion coordinated to His15, which was a mistake by us that it was included in our PDB file; it had been an attempt by us to model the 'shaped' electron density for one coordination site to the Pt bound to the $N^\delta$ of His15, which we had rejected, and was not consistent with our Tanley et al 2014 article. We have considered the preference of Shabalin et al (2015) to model a chlorine in this density and a close by bromine at partial occupancy to explain the 'shape'. However, as the bromide concentration is in huge excess over chloride (by 20 fold), we think that the 4yem Shabalin et al 2015 interpretation highly unlikely, but nevertheless we still cannot offer an explanation for that shape, confirming our earlier analysis described in Tanley et al (2014). The analysis presented here is based on a new diffraction data processing to 1.3 Å resolution. Following Shabalin et al (2015)'s reprocessing of the raw diffraction data for 4g4a, we also redid the diffraction data processing for 4xan to a higher resolution using EVAL (Schreurs et al 2010) concluding in favour of 1.3Å as the resolution limit and which is the basis for our revised PDB file for 4xan (5HMJ). It is very interesting that there is extra X-ray diffraction data from 1.47Å to 1.30 Å resolution e.g. with $<I/\sigma(I)>$ =0.39 and $CC_{1/2}$ = 0.181 in the final shell (1.30 to 1.322Å). In this arXiv article we document in detail our different solvent and split occupancy side chain electron density interpretations as evidence for our statement of approach in our Response article (Tanley et al 2015). Our critical re-examination includes comparisons based on the 4xan diffraction data images reprocessing with three different software packages so as to evaluate the possibility of variations in electron density interpretations due to that. Overall our finalised model (PDB code 5HMJ) (see Table 1) is now improved over 4xan.

## Introduction

A re-refinement of 4xan, hen egg white lysozyme (HEWL) with carboplatin crystallised in NaBr solution, originally at 1.47Å (Tanley et al 2014), has been made (Tanley et al 2016) taking into detailed consideration the valuable Critique of this PDB deposition by Shabalin et al 2015. We have considered their newly refined model (PDB code 4yem) of 4xan in our re-refinement.

## Methods

The diffraction data processing statistics for the 4xan raw diffraction images data reprocessing, now to 1.30 Å, using EVAL (Schreurs et al 2010) are available in Supplementary S1 as Table S1.1. The diffraction data images were also reprocessed using XDS (Kabsch 2010) and MOSFLM (Leslie and Powell 2007).

In order to arrive at the best possible molecular model the method we adopted was as follows. We scrutinised 2Fo-Fc and Fo-Fc electron density maps using COOT (Emsley and Cowtan (2004) calculated using the 1.30 Å EVAL diffraction data. If electron density were absent for any atoms in 4yem, then the XDS and MOSFLM diffraction data electron density maps were also examined. In less clear cases, where we were unsure whether to accept an assignment made in 4yem, we followed the principle "if unsure do not make an assignment".

We have used the highly convenient validation of the protein molecular model incorporated in the refinement program PHENIX_REFINE (Afonine et al 2012), which led to a variety of optimisations of our protein model. Paired refinement assessment of the diffraction resolution (Karplus and Diederichs (2012), Diederichs and Karplus 2013) was made.

## Results

The paired model refinement method (Karplus and Diederichs (2012), Diederichs and Karplus 2013) indicated that the 1.30Å model against the 1.47Å data was better than the refined model against the 1.47Å data (initial and final Rwork and Rfree values were respectively 15.96 vs 16.00 and 21.26 vs 21.31%. Then the 1.47Å refined model was input to the PDB_REDO webserver (Joosten et al 2014) along with our 1.30 Å diffraction data; PDB_REDO concluded as follows:-"*The input model was refined using only data to 1.47Å resolution. Paired refinement was used to establish a new resolution cut-off at 1.30Å. 23087 reflections, 1170 of which were flagged as a test set, were used in the rest of the run. As there were 19.2 reflections per atom available, both an isotropic and an anisotropic B-factor model were considered, and the isotropic B-factor model was selected based on the Hamilton R ratio test.* " As a further check on the resolution limit to finalise on we expanded the resolution limit in a smaller step; PDB REDO similarly concluded that 1.35Å was better than 1.47 Å. As a further, manual, evaluation in yet finer resolution increments we undertook the following checks using the EVAL data in intervals of 0.025 Å starting at 1.30Å and ending at 1.525Å ie using the refined model at the higher resolution with the next increment cut of diffraction data. This indicated that the resolution limit at which the Rfree value began to deteriorate from a best value of 21% around 1.45Å increasing to 22%.

The finalised model deposited at the PDB includes the refined platinum and those ligand atoms at the His15 ND site that could be interpreted namely the two bromines (Figure 1). The Supplementary File 2 provides the details at different stages of the model refinement in finalising our model; this description includes scrutiny of those 'blobs of electron density' that we have not provided an interpretation for and where we explain why in each case.

**Discussion**

On the resolution limit question, which we have revisited compared with our original deposition, which was 1.47Å, we have judged three aspects:-

(i) It is very interesting that there is extra X-ray diffraction data from 1.47Å to 1.30 Å resolution e.g. with $<I/\sigma(I)>$ =0.39 and $CC_{1/2}$ = 0.181 in the final shell (1.30 to 1.322Å). This can presumably be related to the observation, as one watches diffraction images being recorded (JRH), that one does see diffraction spots with clearly visible intensities well beyond where the spot intensities in general have faded into background. As another way of looking at this (Diederichs personal communication) "one can use the one-tailed probability calculation (http://www.danielsoper.com/statcalc3/calc.aspx?id=44):     0.00000074. This means a very small (less than one in a million) probability of $CC_{1/2}$=0.181 to have arisen by chance with 698 reflection pairs, and so this value has a high statistical significance."

(ii) PDB REDO recommends 1.30Å resolution using the 'paired model refinement method' tested both with 1.30Å and with 1.35Å diffraction data and a model at 1.47Å.

(iii) Our manual paired refinement checks indicate that Rfree starts to deteriorate from a best value of 21% around 1.45Å increasing to 22%.

Considering these three points above there is not a perfect consensus on what is the resolution limit. Two of the three points indicate that it is worth preserving the diffraction data to 1.30Å. However, as we do not wish to indicate a 'possible perceived false precision' in knowing the diffraction resolution to two decimal places, we report the resolution limit finally as simply '1.3Å'.

The mistakes we made in solute interpretation that we responded to earlier about (Tanley et al 2015) based on Shabalin et al 2015 have been corrected in our new molecular model. We also removed the four light atoms as a platinum ligand at the His15 ND binding site from the 4xan PDB file that we had mistakenly left in, which was an attempt we made to fit the 'difficult-to-interpret' electron density at that position, but which made our 4xan PDB coordinate file inconsistent with our publication Tanley et al 2014. We have corrected our error. We have considered the preference of Shabalin et al (2015) to model a chlorine at 85% occupancy in this density and a close by bromine at 15% partial occupancy to explain the 'shape'. However, as the bromide concentration is in huge excess over chloride (by 20 fold), we think that the 4yem Shabalin et al 2015 interpretation highly unlikely, but nevertheless we still cannot offer an explanation for that shape, confirming our earlier analysis described in Tanley et al (2014). We have considered, one by one, the additional bound waters in 4yem compared with our 4xan PDB file; these evaluations are described in the Supplementary 2. We have also considered the additional split occupancy amino acid assignments in 4yem; these evaluations are also described in the Supplementary. The Supplementary File 2 lists for these our decisions on each of these bound waters and on each of the split occupancy amino acids as well as solute and ion assignments.

We have two comments to the PDB Validation report section 6.4 on our finalised PDB file, and which are given in Supplementary File S3, regarding our choice of a DMSO solute molecule and an ACT solute molecule to fit their respective electron densities.

## Conclusions

We have considered in considerable detail the four categories of changes made by Shabalin et al 2015 in their PDB file 4yem namely:- (i) their attribution of ligand atoms to a piece of electron density at the bromoplatin binding site, for which we described it as 'difficult to interpret' difference electron density; (ii) their additional bound water molecules; (iii) their additional split occupancy amino acid side chains (iv) their criticisms of some of our solute molecules. Category (iv) has required significant changes to our PDB file 4xan, which we have made in the replacement PDB file (5HMJ). We also removed an organic moiety, which was an attempted fit to a 'shaped piece of electron density' as a ligand in the platinum coordination on the His15 ND side, that we had made but had rejected, as we made clear in our original article and described in our Response article (Tanley et al 2015).

## Acknowledgements
We thank Shabalin et al (2015) for their valuable Critique. We are very grateful to Dr Kay Diederichs for very valuable discussions and provision of XDS processed diffraction data to 1.375Å**.** We thank Dr Colin Levy for helpful comments on the model refinement methodology.

## References
Afonine, P.V. , Grosse-Kunstleve, R.W. , Echols, N., Headd, J.J. , Moriarty, N.W. , Mustyakimov, M. , Terwilliger, T.C. , Urzhumtsev, A. , Zwart, P.H. and Adams. P.D. (2012) *Acta Cryst* D**68**, 352-67 .

Diederichs, K. and Karplus, P. A. (2013) Acta Cryst D**69**, 1215–1222.

Emsley, P. & Cowtan, K. (2004). *Acta Cryst*. D60, 2126–2132.

Joosten, R. P. , Fei Long, Murshudov G. N. , and Perrakis, A. (2014) *IUCrJ* 1, 213-20.

Kabsch, W. *XDS. Acta Cryst.* D**66**, 125-132 (2010).

Karplus, P. A. & Diederichs, K. (2012). Science, 336, 1030–1033.

Leslie, A.G.W. and Powell, H.R. (2007), Evolving Methods for Macromolecular Crystallography, 245, 41-51 ISBN 978-1-4020-6314-5

McNicholas, S. , Potterton, E. , Wilson, K. S. and Noble, M. E. M. (2011) Acta Cryst. D67, 386-394.

Schreurs, A. M. M., Xian, X. & Kroon-Batenburg, L. M. J. (2010). J. Appl. Cryst. 43, 70–82.

Shabalin, I., Dauter, Z., Jaskolski, M., Minor, W. & Wlodawer, A. (2015) *Acta Cryst.* D71, 1965–1979.

Tanley, S. W. M. , Diederichs, K. , Kroon-Batenburg, L. M. J. , Levy, C. , Schreurs, A. M. M. and Helliwell, J. R. Acta Cryst. (2014). F70, 1135-1142.

Tanley, S.W.M., Schreurs, A.M.M., Kroon-Batenburg, L.M.J., Meredith, J., Prendergast, R., Walsh, D., Bryant, P., Levy, C. and Helliwell, J.R. (2012) *Acta Cryst.* D68, 601-612.

Tanley, S. W. M. , Diederichs, K., Kroon- Batenburg, L. M. J. , Levy, C., Schreurs. A. M. M. and Helliwell, J. R. (2015). *Acta Cryst.* D71, 1982–1983.

Tanley, S. W. M. , Kroon-Batenburg, L. M. J. , Schreurs, A. M. M. and Helliwell, J. R. Acta Cryst. (2016) Acta Cryst F72, 253-254.

Figure 1

Stereo views of the electron density at the His15 ND platinum binding site (left hand side of the histidine imidazole ring and oriented with the atoms marked 'x' 'Br 202' top, the Pt is central 'x' and 'Br 204' is at the bottom 'x', and the unexplained electron density at extreme left is labelled '?'; on the NE side the labelled 'x' is 'BR G1' ). (a) the face on view and (b) the edge-on view. This moiety at the His15 ND binding site is clearly square planar in geometry. Blue is the 2Fo-Fc electron density map contoured at 1.2 rms; green is the Fo-Fc electron density map contoured at 5.0σ (the COOT default (Emsley and Cowtan 2004); orange is the anomalous electron density map contoured at 3.0σ.

(a)

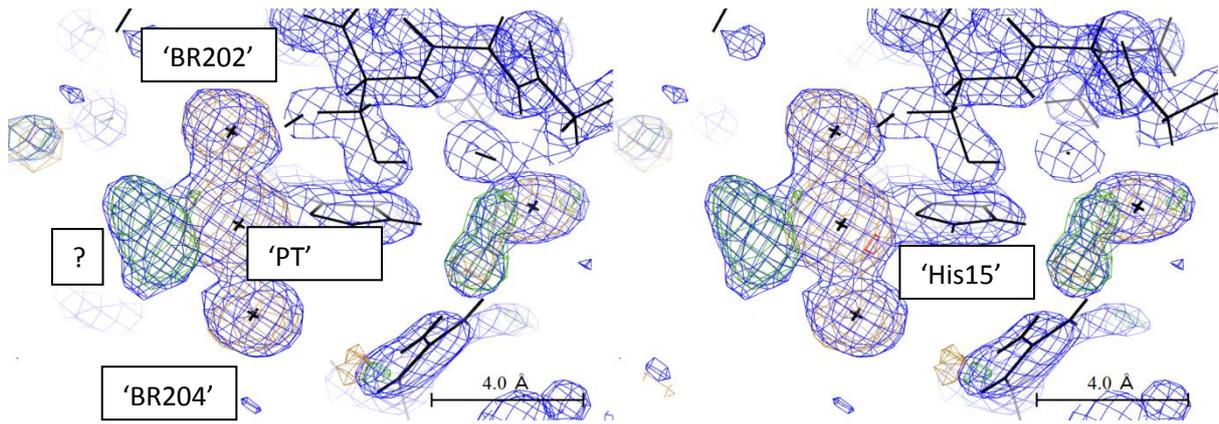

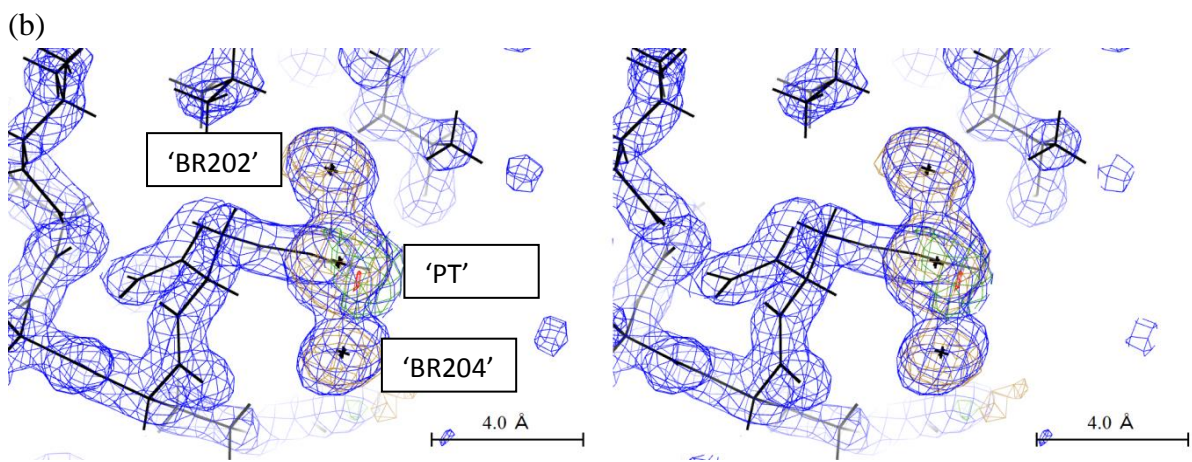

(b)

Table 1 Crystal parameters, data processing and final model refinement statistics.

| PDB code | 5HMJ (replacing 4xan) |
|---|---|
| Wavelength (Å) | 0.9163 |
| Resolution range (Å) | 39.29 - 1.30 (1.346 - 1.30) |
| Space group | P $4_3 2_1 2$ |
| Unit cell | 78.58, 37.29 |
| Total reflections | 250129 (2024) |
| Unique reflections | 23088 (889) |
| Multiplicity | 10.8 (2.3) |
| Completeness (%) | 78.76 (31.05) |
| Mean I/sigma(I) | 10.86 (0.39) |
| Wilson B-factor (Å$^2$) | 14.84 |
| R-meas | 0.095 (2.058) |
| CC$_{1/2}$ | 0.998 (0.181) |
| CC* | 1.000 (0.553) |
| Reflections used for R-free | 1170 (5.34%) |
| R-work | 0.1642 (0.2946) |
| R-free | 0.2150 (0.4248) |
| Cruickshank 'Diffraction Precision Index' (Å) | 0.060 |
| Number of non-hydrogen atoms | 1150 |
|   macromolecules | 1039 |
|   ligands | 29 |
|   water | 82 |
| Protein residues | 129 |
| RMS(bonds) | 0.009 |
| RMS(angles) | 1.18 |
| Ramachandran favored (%) | 98 |
| Ramachandran allowed (%) | 2 |
| Ramachandran outliers (%) | 0 |
| Average atomic B-factor (Å$^2$) | 18.3 |
| B-factor Protein atoms (Å$^2$) | 17.3 |
| B-factor Ligand atoms (Å$^2$) | 31.7 |
| B-factor Solvent atoms (Å$^2$) | 26.0 |

Statistics for the highest-resolution shell are shown in parentheses.

Table 2 Final model refined occupancies and ADP values at the cisplatin binding sites with chemical assignments made on the basis of the omit electron density map of Figure 1.

| Atom | Occupancy* | ADP$^\$$ |
|---|---|---|
| Pt1 | 0.8* | 23 |
| Br1 ('BR202A') | 0.7* | 25 |
| Br2 ('BR204A') | 0.7* | 28 |
| His15ND | 1.0 | 23 |

*Their standard uncertainties are probably about 10% ie 0.1 (ie for an evaluation of the standard uncertainties on a platinum atom in this compound's binding to His15 see Tanley et al 2012).

$These values' standard uncertainties are probably ~ +/- 5 Å$^2$ (Tanley et al 2012).

**Supplementary 1 The EVAL diffraction data processing statistics to 1.30Å.**

```
status
Status for anosad.hkl
RMAT 1 postref
            RMAT                                   DMAT
  -0.0073542   0.0103018   0.0027521     -45.4195213 -64.1266174   0.9385186
  -0.0103831  -0.0073184  -0.0015595      63.6242561 -45.1991425  -9.2219324
   0.0001520  -0.0014932   0.0266309       3.8266640  -2.1683125  37.0279312
Determinant:   0.4342279E-05                          230293.8
cell from rmat: 78.58777 78.58787 37.28824 89.9999 90.0001 89.9997 V= 230293.81
pg constrained: 78.58781 78.58781 37.28824 90.0000 90.0000 90.0000 V= 230293.83
Sigma 0.0491 0.0457 0.0042 0.015 0.014 0.033 Volume 79.93
Bravais=P pg=422
Lambda=0.92 pointgroup=422
reflections=275935 unique=43282 weak=132147 negative=41
Max equivalents 10 one Experiment one Set
presentationscale 0.1708817
Selectable: GOOD WEAK NEGATIVE
Forbid: NONE
Allow: ALL
Require: NONE
total
Rsym=0.087 Rmeas=0.095 Rpim=0.036 Chi2=1.016 nRsym=275123 Unique1=812
Unique2+=42470
<I>=25.813 <s>=3.112 <I/s>=4.59 <I>/<s>=8.294 nMean=275935
rshell
------------------------------------------------------------------------------
```

```
Completeness and Rmerge for Shells
Forbid: NONE
Allow: ALL
Require: NONE
theta from 0.0 to 20.724
Sh Theta  Reso   Meas  Equi   Obs  Mis  Lost  Total  Perc  Cum Uni1 Uni2+ Nrsym Redun  Rsym  Rmeas  Rpim  Chi2
 1  7.49 3.529  17989  3747 21736   192     0  21928  99.1 99.1   12  2722 18711  6.87 0.050 0.054 0.020 1.07
 2  9.45 2.801  18254  3482 21736   246     0  21982  98.9 99.0    0  2722 19440  7.14 0.057 0.061 0.023 1.13
 3 10.84 2.447  15854  5698 21552   408     0  21960  98.1 98.7   27  2670 17032  6.38 0.068 0.074 0.029 1.14
 4 11.94 2.223  17382  4374 21756   218     0  21974  99.0 98.8    5  2717 18856  6.94 0.081 0.088 0.033 1.18
 5 12.88 2.064  17366  4142 21508   364     0  21872  98.3 98.7   10  2681 18989  7.08 0.094 0.101 0.038 1.16
 6 13.70 1.942  17544  4032 21576   490     0  22066  97.8 98.5   14  2685 19254  7.17 0.116 0.125 0.046 1.14
 7 14.44 1.845  15081  6391 21472   522     0  21994  97.6 98.4   54  2632 16579  6.30 0.147 0.161 0.064 1.12
 8 15.11 1.764  15953  5463 21416   434     0  21850  98.0 98.4   22  2657 17691  6.66 0.192 0.208 0.080 1.10
 9 15.73 1.696  16395  5013 21408   488     0  21896  97.8 98.3   29  2648 18237  6.89 0.245 0.265 0.100 1.06
10 16.31 1.638  16427  4637 21064   962     0  22026  95.6 98.0    3  2631 18396  6.99 0.295 0.318 0.119 1.02
11 16.85 1.587  16276  4508 20784  1338     0  22122  94.0 97.7   33  2567 18164  7.08 0.371 0.400 0.149 0.98
12 17.37 1.541  15518  4356 19874  1962     0  21836  91.0 97.1   42  2444 17386  7.11 0.508 0.548 0.204 0.94
13 17.85 1.501  13209  5513 18722  3048     0  21770  86.0 96.3   60  2280 14747  6.47 0.597 0.650 0.253 0.84
14 18.31 1.464   9733  5611 15344  6678     0  22022  69.7 94.4   75  1843 10953  5.94 0.707 0.777 0.313 0.81
15 18.75 1.431   7710  5450 13160  8802     0  21962  59.9 92.1   56  1589  8722  5.49 0.785 0.871 0.365 0.76
16 19.18 1.400   6057  5239 11296 11052     0  22348  50.5 89.4   73  1339  6849  5.12 1.005 1.125 0.488 0.66
17 19.59 1.372   4679  4713  9392 12138     0  21530  43.6 86.8   91  1083  5266  4.86 1.111 1.254 0.562 0.64
18 19.98 1.346   3887  4713  8600 13664     0  22264  38.6 84.1   54  1021  4435  4.34 1.377 1.575 0.742 0.58
19 20.36 1.322   2791  4425  7216 14546     0  21762  33.2 81.4   61   841  3175  3.78 1.481 1.741 0.891 0.61
20 20.72 1.300   2024  4284  6308 15752     0  22060  28.6 78.8   91   698  2241  3.21 1.700 2.058 1.137 0.60
===============================================================================================================
   20.72 1.300 250129 95791 345920 93304     0 439224 78.8 78.8  812 42470 275123 6.48 0.087 0.095 0.036 1.02
Resolution 39.294-1.3 (1.322-1.3)
Rsym 0.087 (1.7) Rmeas 0.095 (2.058) Rpim 0.036 (1.137)
Unique 43282 (789) Completeness 78.757 (28.595)
list intensity
-------------------------------------------------------------------------------
Intensity distribution for Shells, unmerged and merged
-------------------------------------------------------------------------------
Intensity distribution for Shells, unmerged and merged
Forbid: NONE
Allow: ALL
Require: NONE
Sh Theta  Reso      N     <I>   <s> <I/s>  Nmerge   <I>   <s> <I/s>  cc1/2   cc*  npair
 1  7.49 3.529  18723 138.87  9.77 12.60    2734 140.51 4.06 31.70 0.997 0.999  2722
 2  9.45 2.801  19440  86.15  6.62 10.82    2722  86.72 2.62 27.96 0.998 0.999  2722
 3 10.84 2.447  17059  44.19  3.99  9.03    2697  44.30 1.76 21.67 0.995 0.999  2670
 4 11.94 2.223  18861  32.21  3.34  7.73    2722  32.48 1.36 19.68 0.994 0.999  2717
 5 12.88 2.064  18999  24.25  2.91  6.59    2691  24.28 1.15 17.00 0.994 0.999  2681
 6 13.70 1.942  19268  16.68  2.48  5.35    2699  16.77 0.97 13.97 0.993 0.998  2685
 7 14.44 1.845  16633  10.95  2.12  4.07    2686  11.29 0.98  9.77 0.985 0.996  2632
 8 15.11 1.764  17708   7.84  1.98  3.19    2679   7.99 0.82  8.04 0.980 0.995  2657
 9 15.73 1.696  18271   5.77  1.89  2.47    2677   5.90 0.76  6.35 0.973 0.993  2648
10 16.31 1.638  18399   4.61  1.86  2.01    2634   4.68 0.73  5.23 0.965 0.991  2631
11 16.85 1.587  18197   3.62  1.87  1.60    2600   3.68 0.74  4.15 0.949 0.987  2567
12 17.37 1.541  17422   2.57  1.86  1.16    2486   2.57 0.73  2.99 0.911 0.976  2444
13 17.85 1.501  14813   2.07  1.87  0.92    2340   2.01 0.82  2.22 0.867 0.964  2280
14 18.31 1.464  11028   1.69  1.86  0.77    1918   1.61 0.89  1.73 0.771 0.933  1843
15 18.75 1.431   8778   1.50  1.88  0.68    1645   1.39 0.94  1.47 0.690 0.904  1589
16 19.18 1.400   6922   1.08  1.89  0.48    1412   1.03 0.99  1.01 0.556 0.845  1339
17 19.59 1.372   5357   0.98  1.92  0.42    1174   0.95 1.06  0.86 0.567 0.851  1083
18 19.98 1.346   4489   0.75  1.93  0.32    1075   0.65 1.07  0.61 0.297 0.677  1021
19 20.36 1.322   3236   0.72  1.97  0.29     902   0.68 1.18  0.54 0.268 0.650   841
20 20.72 1.300   2332   0.60  2.01  0.23     789   0.56 1.31  0.39 0.181 0.553   698
===============================================================================================
   20.72 1.300 275935  25.81  3.11  4.59   43282  24.23 1.29 10.86 0.998 1.000 42470
r
```

**Supplementary 2 The stages involved in arriving at the final re refined molecular model**

Here below we provide the detailed list for those key issues mentioned in the main text in finalising our model. At this stage of the model refinement we have used the three independently processed diffraction data, namely from EVAL (to 1.30Å), XDS (to 1.375Å) and MOSFLM (to 1.35Å). Table S2.1 presents the diffraction data processing and model refinement statistics for each.

**Table S2.1 Crystal parameters, data processing and final model refinement statistics for EVAL (to 1.30Å), XDS (to 1.375Å) and MOSFLM (to 1.35Å).**

|  | EVAL | XDS | MOSFLM |
|---|---|---|---|
| **Unit cell parameters** | 78.58, 37.29 | 78.61, 37.29 | 78.58, 37.29 |
| **Wavelength (Å)** | 0.9163 | 0.9163 | 0.9163 |
| **Resolution range (Å)** | 39.29 - 1.3 (1.346 - 1.3) | 39.29 - 1.372 (1.421 - 1.372) | 35.14 - 1.35 (1.398 - 1.35) |
| **Unique reflections** | 23088 (889) | 21373 (811) | 21976 (1012) |
| **Completeness (%)** | 78.76 (31.05) | 85.4 (40.9) | 83.77 (39.39) |
| **Mean I/sigma(I)** | 10.86 (0.39) | 12.11 (1.50) | 16.0 (2.1) |
| **Wilson B-factor** | 14.84 | 12.37 | 10.65 |
| **R-meas** | 0.095 (2.058) | 0.097 (0.847) | 0.119 (1.005) |
| **$CC_{1/2}$** | 0.998 (0.181) | 0.997 (0.592) | 0.998 (0.60) |
| **CC*** | 1.000 (0.553) |  | N/A |
| **Reflections used for R-free** | 1170 (5.34%) | 1060 (4.96%) | 1107 (5.04%) |
| **R-work** | 0.1658 (0.2932) | 0.1524 (0.2348) | 0.1543 (0.2506) |
| **R-free** | 0.2165 (0.4304) | 0.1925 (0.2650) | 0.1933 (0.2849) |
| **Number of non-hydrogen atoms** | 1139 | 1139 | 1139 |
| **Cruickshank 'dpi' (Å)** | 0.060 | 0.061 | 0.059 |
| macromolecules | 1039 | 1039 | 1039 |
| ligands | 22 | 22 | 22 |
| water | 78 | 78 | 78 |
| **Protein residues** | 0 | 0 | 0 |
| **RMS(bonds)** | 0.009 | 0.008 | 0.008 |
| **RMS(angles)** | 1.19 | 1.17 | 1.18 |
| **Ramachandran favored (%)** | 98 | 98 | 98 |
| **Ramachandran allowed (%)** | 2 | 2 | 2 |
| **Ramachandran outliers (%)** | 0 | 0 | 0 |
| **Clashscore** | 0.97 | 0.97 | 0.97 |
| **Average atomic B-factor (Å$^2$)** | 18.10 | 15.20 | 13.90 |
| **B-factor Protein atoms (Å$^2$)** | 17.30 | 14.40 | 13.10 |
| **B-factor Ligand atoms (Å$^2$)** | 30.90 | 28.30 | 27.00 |
| **B-factor Solvent atoms (Å$^2$)** | 25.20 | 22.30 | 21.00 |

**Part A; The bound water structure**

Method adopted:- We scrolled one by one through the 4yem bound water assignments and saw how they matched the 1.30 Angstrom EVAL electron density. If electron density was absent we then consulted the 1.37 Å XDS and 1.35Å resolution MOSFLM electron densities at each respective position to make a final decision. In general, where we were unsure whether to accept an assignment made in 4yem we followed the principle "if unsure do not make an assignment". This statement is made in each case below that needed that particular decision. In the figures below each of the three (2Fo-Fc) maps are set at 1.2rms, the (Fo-Fc) map is set at COOT's default of 5.0 sigma and the EVAL anomalous difference density map, calculated at 2.0 Angstrom resolution, is contoured at 3.0 sigma. Colour scheme; EVAL 2Fo-Fc blue, Fo-Fc green, anomalous difference Fourier map orange; XDS 2Fo-Fc turqoise; MOSFLM 2Fo-Fc mauve.

W302; no electron density evidence for this in any of EVAL, XDS or MOSFLM maps. Assignment of 4yem not accepted.

W304; tiny electron density evidence (1.2 rms 2Fo-Fc blips) for this in XDS and MOSFLM maps. Assignment of 4yem not accepted.

W305; no electron density evidence for this in any of EVAL, XDS or MOSFLM maps. Assignment of 4yem not accepted.

W306; no electron density evidence for this in any of EVAL, XDS or MOSFLM maps. Assignment of 4yem not accepted.

W307; no electron density evidence for this in any of EVAL, XDS or MOSFLM maps. Assignment of 4yem not accepted.

W308; tiny electron density evidence (1.2 rms 2Fo-Fc blip) for this in EVAL map. Assignment of 4yem not accepted.

W309; clear evidence in all three of our 2Fo-Fc maps. Added to our model.

W310; clear evidence in all three of our 2Fo-Fc maps. Added to our model.

W311; clear evidence in all three of our 2Fo-Fc maps. Added to our model.

W312; the map is too shaped and with an anom peak very close to be a Water. But unable to assign anything with confidence; so again follow the principle "if unsure do not make an assignment".

W313; clear evidence in all three of our 2Fo-Fc maps. Added to our model.

W314; no electron density evidence for this in any of EVAL, XDS or MOSFLM maps. Assignment of 4yem not accepted.

W315; no electron density evidence for this in any of EVAL, XDS or MOSFLM maps. Assignment of 4yem not accepted.

W317; not nicely spherical but not too aspherical and in EVAL, XDS and MOSFLM 2Fo-Fc and Fo-Fc maps. Added to our model.

W318; 4yem has this as a 0.7 occupied water as part of the interpretation of the NE side Pt ligands. There is some misshapen density but is only 2.1 Angstrom from the Arg14 nitrogen. Cannot be assigned with confidence, so do not accept this.

W321; weak 'blip' level of evidence in each 2Fo-Fc map but not in their Fo-Fc maps. so do not accept this.

W322; no electron density evidence for this in any of EVAL, XDS or MOSFLM maps. Assignment of 4yem not accepted.

W324; no electron density evidence for this in any of EVAL, XDS or MOSFLM maps. Assignment of 4yem not accepted.

W325; no electron density evidence for this in any of EVAL, XDS or MOSFLM maps. Assignment of 4yem not accepted.

W327; no electron density evidence for this in any of EVAL, XDS or MOSFLM maps. Assignment of 4yem not accepted.

W328; clear evidence in all three of our 2Fo-Fc maps. Added to our model.

W334; a 0.3 occupancy water. Weak 2Fo-Fc map evidence but in all three maps. Borderline case. But again follow the principle "if unsure do not make an assignment".

W336; no electron density evidence for this in any of EVAL, XDS or MOSFLM maps. Assignment of 4yem not accepted.

W340; clear evidence in all three of our 2Fo-Fc maps. Added to our model.

W342; no electron density evidence for this in any of EVAL, XDS or MOSFLM maps. Assignment of 4yem not accepted.

W347; marginal, but consistent, evidence in all three of our 2Fo-Fc maps. Added to our model.

W348; clear evidence in all three of our 2Fo-Fc maps. Added to our model.

W350; no electron density evidence for this in any of EVAL, XDS or MOSFLM maps. Assignment of 4yem not accepted.

W351; no electron density evidence for this in any of EVAL, XDS or MOSFLM maps. Assignment of 4yem not accepted.

W354; no electron density evidence for this in any of EVAL, XDS or MOSFLM maps. Assignment of 4yem not accepted.

W355; clear evidence in all three of our 2Fo-Fc maps. Added to our model.

W356; one 'blip' only (EVAL 2Fo-Fc) and no other evidence. Assignment of 4yem not accepted.

W358; a high B water (43) but with small peaks in all three of our maps slightly off position Marginal case. But again follow the principle "if unsure do not make an assignment".

W359; no electron density evidence for this in any of EVAL, XDS or MOSFLM maps. Assignment of 4yem not accepted.

W361; too much shape to be a water, also a 3.3 sigma anom peak, but what is it? But again follow the principle "if unsure do not make an assignment":-

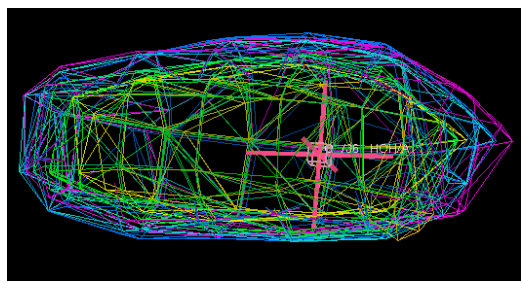

W362; marginal evidence but in all three of our 2Fo-Fc maps. Added to our model.

W363; a 0.5 occupancy water in 4yem with low B factor (15) and high peak height in our 2Fo-Fc map eg EVAL 3.3rms, nicely spherical. We have a 2.7σ anomalous map peak. Best option a Br.

W364; no electron density evidence for this in any of EVAL, XDS or MOSFLM maps. Assignment of 4yem not accepted.

W365; clear evidence in all three of our 2Fo-Fc maps. Added to our model.

W368; no electron density evidence for this in any of EVAL, XDS or MOSFLM maps. Assignment of 4yem not accepted.

W370; no electron density evidence for this in any of EVAL, XDS or MOSFLM maps. Assignment of 4yem not accepted.

W371; Similar case to W363; but here a full occupancy water but with low B factor (17) and high peak height in the 2Fo-Fc map eg EVAL 3.7rms, nicely spherical. We have a 2.9σ anomalous map peak. Best option a Br.

W372; clear evidence in all three of our 2Fo-Fc maps. Added to our model.

W374; no electron density evidence for this in any of EVAL, XDS or MOSFLM maps. Assignment of 4yem not accepted.

W375; no electron density evidence for this in any of EVAL, XDS or MOSFLM maps. Assignment of 4yem not accepted.

W376; no electron density evidence for this in any of EVAL, XDS or MOSFLM maps. Assignment of 4yem not accepted.

W377; no electron density evidence for this in any of EVAL, XDS or MOSFLM maps. Assignment of 4yem not accepted.

W378; no electron density evidence for this in any of EVAL, XDS or MOSFLM maps. Assignment of 4yem not accepted.

W379; evidence is 'small peaks' and in only two of the 2Fo-Fc maps (MOSFLM and XDS). Assignment not accepted.

W380; no electron density evidence for this in any of EVAL, XDS or MOSFLM maps. Assignment of 4yem not accepted.

W381; too much shape in each 2Fo-Fc map to be a water, (and there is a 3.0σ anom peak, albeit a 'small blip') but what is it? So again follow the principle "if unsure do not make an assignment" :-

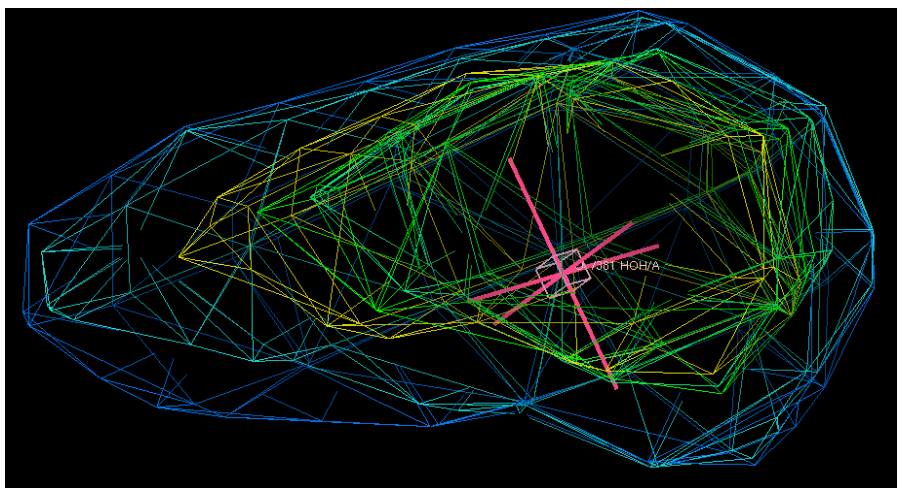

W383; no electron density evidence for this in any of EVAL, XDS or MOSFLM maps. Assignment of 4yem not accepted.

W385; evidence in all three of our 2Fo-Fc maps. Added to our model.

W386; no electron density evidence for this in any of EVAL, XDS or MOSFLM maps. Assignment of 4yem not accepted.

W387; evidence in all three of our 2Fo-Fc maps. Added to our model.

W388; no electron density evidence for this in any of EVAL, XDS or MOSFLM maps. Assignment of 4yem not accepted.

W389; no electron density evidence for this in any of EVAL, XDS or MOSFLM maps. Assignment of 4yem not accepted.

W391; no electron density evidence for this in any of EVAL, XDS or MOSFLM maps. Assignment of 4yem not accepted.

W393; marginal evidence but in all three of our 2Fo-Fc maps. Added to our model.

W394; evidence in all three of our 2Fo-Fc maps and all three of the Fo-Fc maps. Added to our model.

W395; too much shape in each 2Fo-Fc map to be a water, (and no anom peak) but what is it? So again follow the principle "if unsure do not make an assignment":-

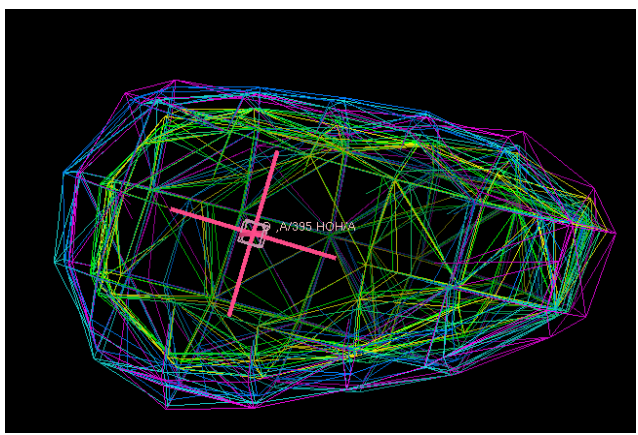

W396; no electron density evidence for this in any of EVAL, XDS or MOSFLM maps. Assignment of 4yem not accepted.

W397; no electron density evidence for this in any of EVAL, XDS or MOSFLM maps. Assignment of 4yem not accepted.

W399; no electron density evidence for this in any of EVAL, XDS or MOSFLM maps. Assignment of 4yem not accepted.

W400; no electron density evidence for this in any of EVAL, XDS or MOSFLM maps. Assignment of 4yem not accepted.

W401; no electron density evidence for this in any of EVAL, XDS or MOSFLM maps. Assignment of 4yem not accepted.

W403; no electron density evidence for this in any of EVAL, XDS or MOSFLM maps. Assignment of 4yem not accepted.

W404; evidence in all three of our 2Fo-Fc maps. Added to our model.

W405; no electron density evidence for this in any of EVAL, XDS or MOSFLM maps. Assignment of 4yem not accepted.

W406; Clear spherical peaks in all three 2Fo-Fc maps and all three Fo-Fc maps and a 4 sigma anom peak. 4yem has this as a 0.5 occupied water and has a B of 19. If it is a bromide ion what is it interacting with? There are two peptide NHs (both Arg45, one a symmetry mate) but each rather too far away:-

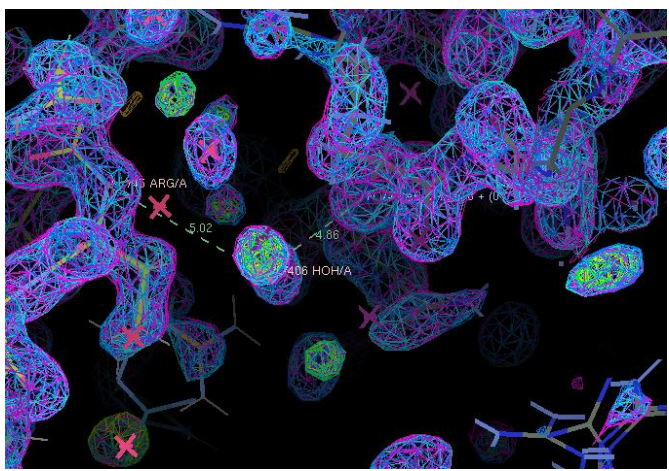

This is a marginal decision but given the firm map evidence and two, albeit far away, NHs assigned this to be a bromide ion.

W407; no electron density evidence for this in any of EVAL, XDS or MOSFLM maps. Assignment of 4yem not accepted.

W408; no electron density evidence for this in any of EVAL, XDS or MOSFLM maps. Assignment of 4yem not accepted.

W409; evidence in all three of our 2Fo-Fc maps and our Fo-Fc maps. Added to our model.

W410 no electron density evidence for this in any of EVAL, XDS or MOSFLM maps. Assignment of 4yem not accepted.

W441; evidence in all three of our 2Fo-Fc and our Fo-Fc maps. Added to our model.

W443; evidence in all three of our 2Fo-Fc maps. Added to our model.

W444 weak electron density evidence for this in EVAL and XDS maps, also not in the MOSFLM map. Assignment of 4yem not accepted.

W446 no electron density evidence for this in any of EVAL, XDS or MOSFLM maps. Assignment of 4yem not accepted.

W447 no electron density evidence for this in any of EVAL, XDS or MOSFLM maps. Assignment of 4yem not accepted.

W448 no electron density evidence for this in any of EVAL, XDS or MOSFLM maps. Assignment of 4yem not accepted.

W449 no electron density evidence for this in any of EVAL, XDS or MOSFLM maps. Assignment of 4yem not accepted.

W450 no electron density evidence for this in any of EVAL, XDS or MOSFLM maps. Assignment of 4yem not accepted.

W451 no electron density evidence for this in any of EVAL, XDS or MOSFLM maps. Assignment of 4yem not accepted.

W452 no electron density evidence for this in any of EVAL, XDS or MOSFLM maps. Assignment of 4yem not accepted.

W453 no electron density evidence for this in any of EVAL, XDS or MOSFLM maps. Assignment of 4yem not accepted.

W454 no electron density evidence for this in any of EVAL, XDS or MOSFLM maps. Assignment of 4yem not accepted.

W456; evidence in all three of our 2Fo-Fc maps and in the MOSFLM Fo-Fc map. Added to our model.

W457 weak electron density evidence for this in XDS or MOSFLM maps but not in EVAL 2Fo-Fc map. Assignment of 4yem not accepted.

W458; evidence in all 2Fo-Fc and Fo-Fc maps and a 4.0 sigma anom peak. Therefore surely not a bound water! But if a bromide ion what is it interacting with? There is nothing obvious, So again follow the principle "if unsure do not make an assignment".

W459; evidence in all three of our 2Fo-Fc maps and in the MOSFLM Fo-Fc map. Added to our model.

W460 no electron density evidence for this in any of EVAL, XDS or MOSFLM maps. Assignment of 4yem not accepted.

W461 no electron density evidence for this in any of EVAL, XDS or MOSFLM maps. Assignment of 4yem not accepted.

W462 no electron density evidence for this in any of EVAL, XDS or MOSFLM maps. Assignment of 4yem not accepted.

W463 no electron density evidence for this in any of EVAL, XDS or MOSFLM maps. Assignment of 4yem not accepted.

W464 no electron density evidence for this in any of EVAL, XDS or MOSFLM maps. Assignment of 4yem not accepted.

W465 no electron density evidence for this in any of EVAL, XDS or MOSFLM maps. Assignment of 4yem not accepted.

W466 no electron density evidence for this in any of EVAL, XDS or MOSFLM maps. Assignment of 4yem not accepted.

W467 no electron density evidence for this in any of EVAL, XDS or MOSFLM maps. Assignment of 4yem not accepted.

W468 no electron density evidence for this in any of EVAL, XDS or MOSFLM maps. Assignment of 4yem not accepted.

W469 no electron density evidence for this in any of EVAL, XDS or MOSFLM maps. Assignment of 4yem not accepted.

W471; this is a very interesting and complex interpretation in 4yem. The core of the interpretation is a split occupancy, 40% and 60%, for the Asp 18 side chain. This leads 4yem to assign a 40% Water 471 and a 60 % Water 320. Ie the idea presumably being that for the 40% of the side chain case the 60% Water 320 is present. In our EVAL model we have no split occupancy for Asp18, for which there is no evidence anyway, and a fully occupied Water 320 with a sensible B of 21. Assignments of 4yem not accepted.

W472 no electron density evidence for this in any of EVAL, XDS or MOSFLM maps. Assignment of 4yem not accepted.

W473 no electron density evidence for this in any of EVAL, XDS or MOSFLM maps. Assignment of 4yem not accepted.

W474 A small blip in the 2Fo-Fc MOSFLM map but otherwise no electron density evidence for this water in the EVAL or XDS maps. Assignment of 4yem not accepted.

W475 a 50% occupied, low B water (15); I have this as an Arg61 NH1. The continuity of density to Arg 61 CZ admittedly is poor including small negative peaks. But to reassign to a bound water? Again follow the principle "if unsure do not make an assignment". However in this case what to do about our placement of Arg 61 NH1? Left 'as is'.

W476 no electron density evidence for this in any of EVAL, XDS or MOSFLM maps. Assignment of 4yem not accepted.

W477 no electron density evidence for this in any of EVAL, XDS or MOSFLM maps. Assignment of 4yem not accepted.

W478 A small blip in the 2Fo-Fc MOSFLM map but otherwise no electron density evidence for this water in the EVAL or XDS maps. Assignment of 4yem not accepted.

W479 no electron density evidence for this in any of EVAL, XDS or MOSFLM maps. Assignment of 4yem not accepted.

W480 no electron density evidence for this in any of EVAL, XDS or MOSFLM maps. Assignment of 4yem not accepted.

**We now revisit all our solvent/solute assignments in our current (we list only differences in interpretation between our model and 4yem ie no mention means they agree in their assignments):-**

W311 is in our model but not 4yem. 2Fo-Fc map evidence looks very good. Keep W311.

W316 is in our model but not 4yem. 2Fo-Fc map evidence looks very good. Keep W316.

W322 is in our model but not 4yem. 2Fo-Fc map evidence looks very good. Keep W322.

W375 has a 3.0 sigma blip anom peak but we have a Water and so does 4yem. Leave the assignment W375 being a water as is.

W386 electron density has shape. 4yem has this as a water too (W470). Unsure about the assignment but since too much shape to leave as a water, delete it from our model. Maybe a DMSO?

Br B has some shape but a clear anom peak. Leave as a bromide ion.

WK12 has all maps with density even though not assigned in 4yem. Retain as WK12.

WK19 has all maps with density even though not assigned in 4yem. Retain as WK19.

**We now revisit the new waters added from the 4yem comparison above:-**

Delete waters 'K' 9, 10, 11, 13, 14, 16, 17, 21, 23, 28 deleted: (each had moved out of their 2Fo-Fc electron density)

Revisit the 300 upwards numbered waters:-

Delete W311, 313, 321,322, 341: (each had moved out of their 2Fo-Fc electron density).

Delete W382, 385: (marginal electron density).

**Part B Amino acid split occupancy assignments assessed against all three of our EVAL, XDS and MOSFLM diffraction data processed electron density maps.**

See our comments made in the table below.

| 4yem assignments:- | 4xan revision needed?:- |
| --- | --- |
| **Double occupancy side chains instead of single occupancy:-** | |
| Lys-1 | No evidence for split occupancy either in Fo-Fc or 2Fo-Fc maps |

| Asp-18 | This is a very interesting and complex interpretation by 4yem. The core of the interpretation is a split occupancy 40% and 60% for Asp 18 side chain. This leads 4yem to assign a 40% water 471 and a 60 % Water 320. Ie the idea being that for the 40% of the side chain case the 60% water 320 is present. In our EVAL model we have no split occupancy for Asp18, for which there is no evidence anyway, and a fully occupied Water 320 with a sensible B of 21. Assignments of 4yem not accepted. |
|---|---|
| Asn-19 | There is evidence for a split occupancy side chain and added to our model. |
| Arg-21 | There is evidence for a split occupancy side chain and added to our model. |
| Arg-45 | After several refinement cycles and checks and rechecks it is still not terribly clear what the Arg 45 and Arg 68 side chains are doing. We have opted for, and PHENIX Refine has refined, the relative occupancies of Arg 68 split 70%/30% and Arg 45 single ie full occupancy. |
| Arg-68 | There is evidence for a split occupancy side chain and added to our model. |
| Ile-55 | There is evidence of a split occupancy but none of the COOT alternative rotamers makes a fit Still leave as 100% single occupancy. |
| Arg-61 | No evidence of a split occupancy unless one goes down to (2Fo-Fc) 0.4rms or so, which we don't think is credible to go that low in the contouring. Assignment of a split occupancy not accepted. |
| Asn-77 | There is evidence for a split occupancy side chain and added to our model. |
| Asn-93 | No electron density evidence for a |
|   | split occupancy. |
| Asp-101 | No electron density evidence for a split occupancy. |

**Part C  We now list below our checks, one by one, of those solute and bound water assignments of Shabalin et al 2015, ie in their PDB file 4yem, including showing our EVAL, XDS and MOSFLm electron density maps in detail.**

| 4yem's Anions/Cations/DMSOs:- | Our decisions :- |
|---|---|
| Swapped DMS 212 for two partially, each low percentage, occupied bromines close together | DMS212 deleted.  4yem assignment not accepted. |
| Swapped DMS 216 for Br 212 | DMS216 deleted. Assignment of 4yem not accepted due to aspherical shape of the density. |
| Br-209 | Assignment not accepted due to aspherical shape of the density. |
| Br-214 | Assignment not accepted due to aspherical shape of the density. |
| Br-215 (split occupancy 0.6 and 0.4) | Accepted the major Br peak. |
| ACT 222 (reassigned from a DMS in 4xan) | Accepted and added to our model. |

Br-204 15%, along with Cl-223 85%, in 4yem ie the cisplatin His 15 ND side 'extreme left hand, not explainable by us, electron density'. No anomalous difference density peak until one goes down to 2.2 sigma, and then it is in the wrong place ie we believe this is 'noise'. Our view then is that the 4yem interpretation is not clear ie an assignment is not possible within these diffraction data and electron density maps.

Br212; the density is elongated:-

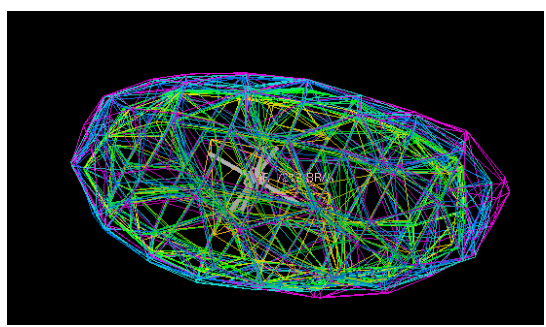

A different assignment seems likely. The anomalous map peak is 3.7 sigma so presumably too large for a DMSO. But how to interpret this? Leave unassigned.

Br214; 4yem has as a 0.25 occupied Br. There is extended density, including anomalous, interpreted by 4yem as Pt218 at 0.1 occupancy liganded to Lys96 and a fully occupied bound water 301:-

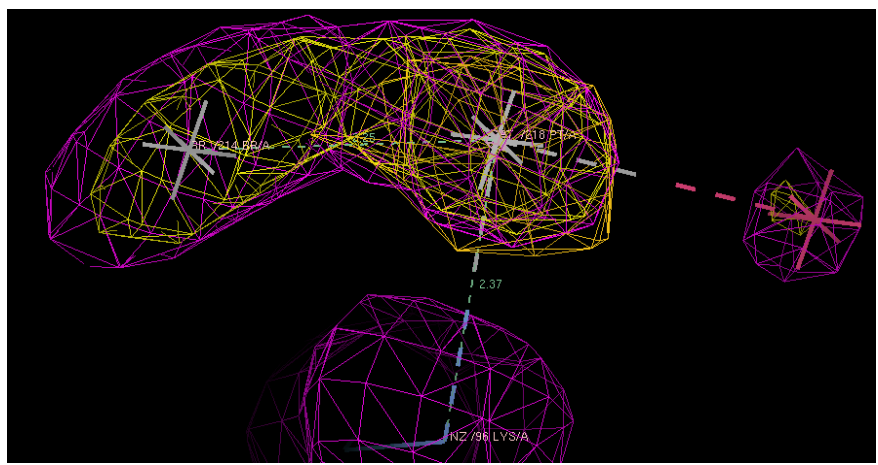

This might be a plausible assignment. Unsure whether to accept this; so follow the principle "if unsure do not make an assignment".

Cl223; the left hand density of the His15ND side. No anom peak. No obvious reason to assign to a chloride ion; Shabalin et al themselves declare this an 'audacious' assignment justified because of the long crystal storage shelf time (a year) and thereby exposure to chloride. Unsure so followed the principle "if unsure do not make an assignment".

Pt224; a 0.15 occupancy Pt. There is a 4.4 sigma anom peak here and in 4xan we also made this assignment and very similar occupancy (0.13). Thus far this has been omitted for this revisit to have an omit map here. We now conclude that we are unsure whether to finally assign this to a low occupancy Pt in our deposited revisit coordinates file; it seems implausible. No assignment made.

Br1H deleted (ADP=233Å$^2$).

**Part D Our cycle 8 difference map and EVAL Cycle 8 model so as to evaluate those remaining difference Fourier map peaks and actions taken by us:-**

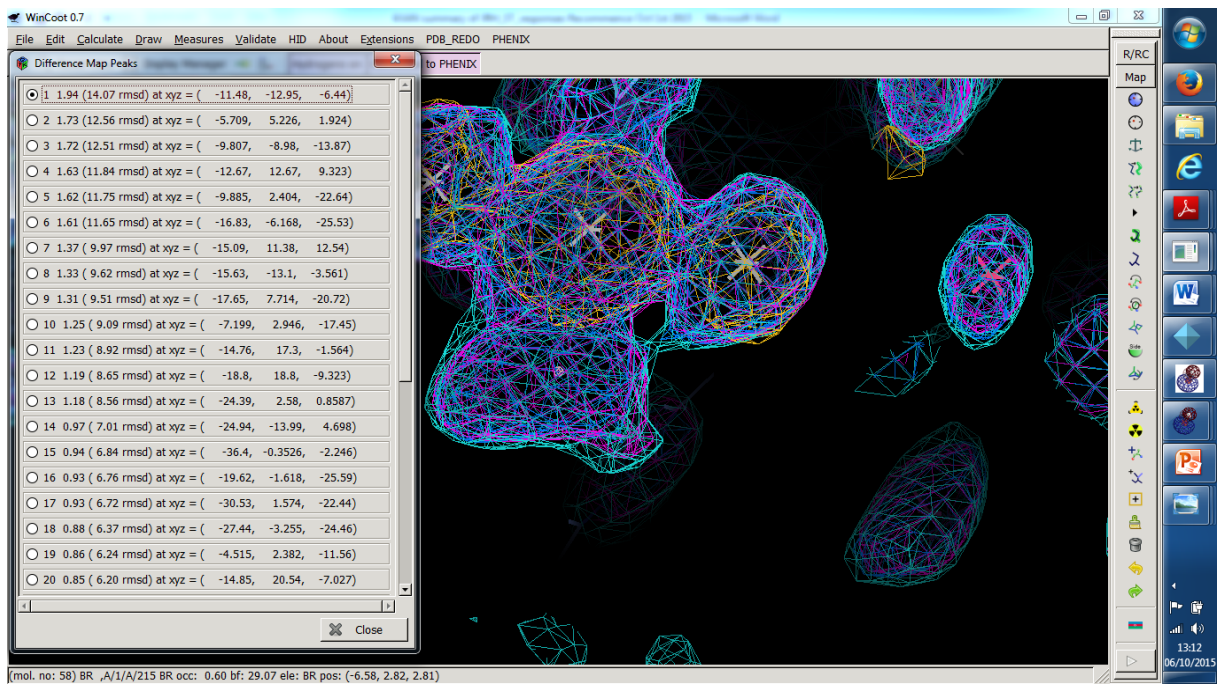

Peak 1:-

Still unclear what this is! As we stated in our paper Tanley et al 2014 originally.

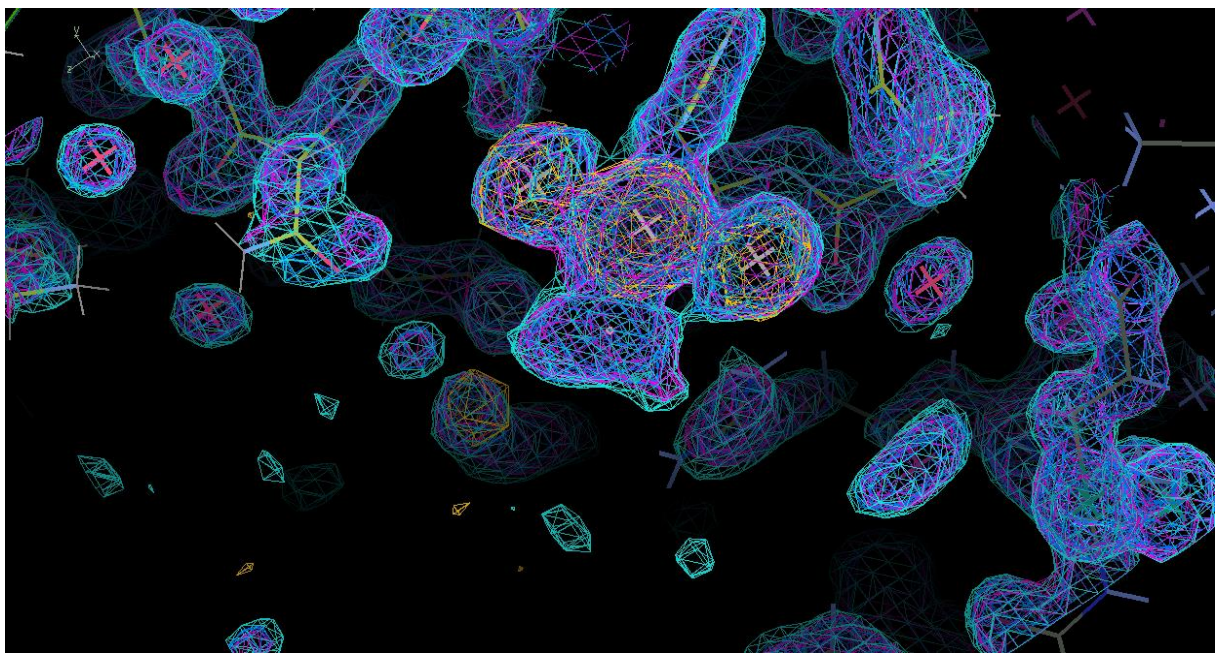

Peak 2:-

This is a minor occupancy Br in 4yem, but what is the structural chemistry reason for this?

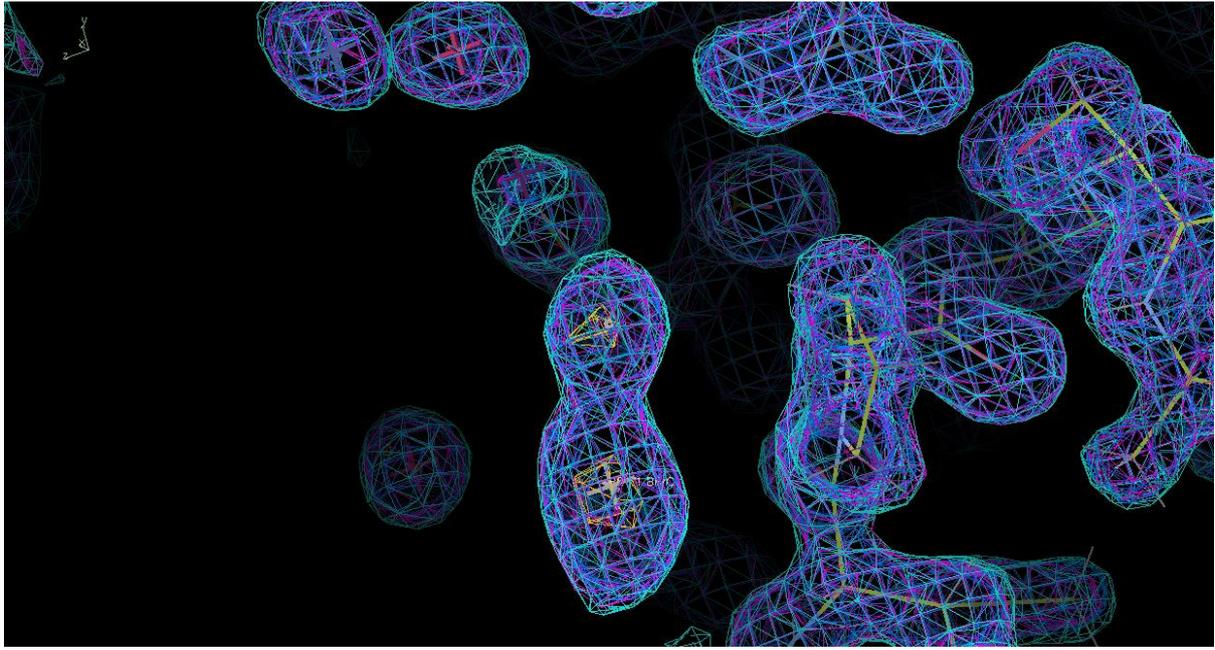

Peak 3:-

This is assigned as a low occupancy Pt at the His 15 NE side in both 4xan (13%) and in 4yem (15%). Is that really plausible? Also by keeping it omitted throughout the new EVAL refinement a further peak in the 2Fo-Fc and Fo-Fc maps has appeared (with no anom peak). What is that? Again follow the principle "if unsure do not make an assignment".

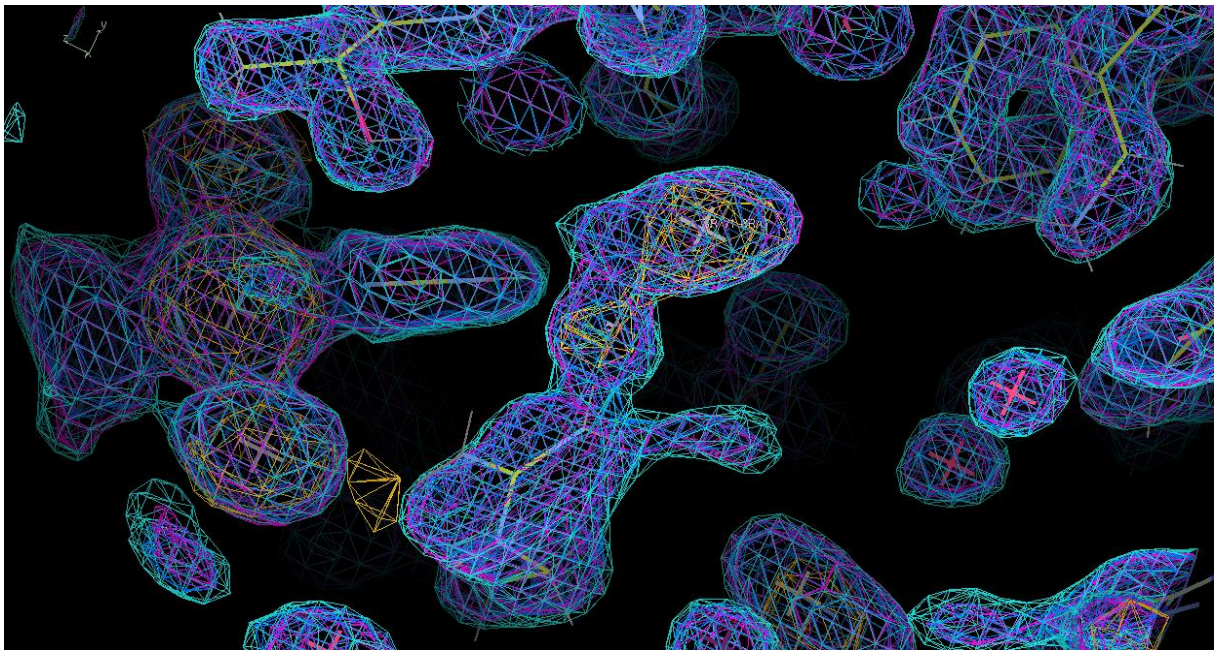

Peak 4:-

Now assigned as a Cl due to the review above of all three sets of maps (EVAL, XDS and MOSFLM).

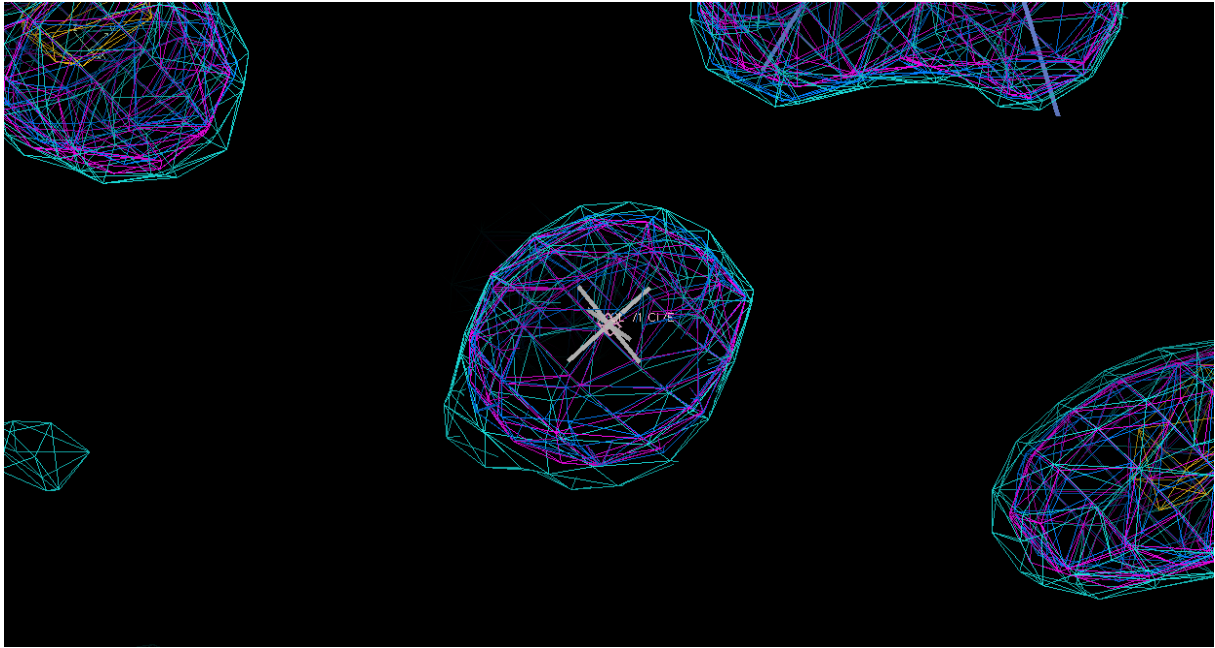

Peak 5:-

Now assigned as a Cl due to the review above of all three sets of maps (EVAL, XDS and MOSFLM).

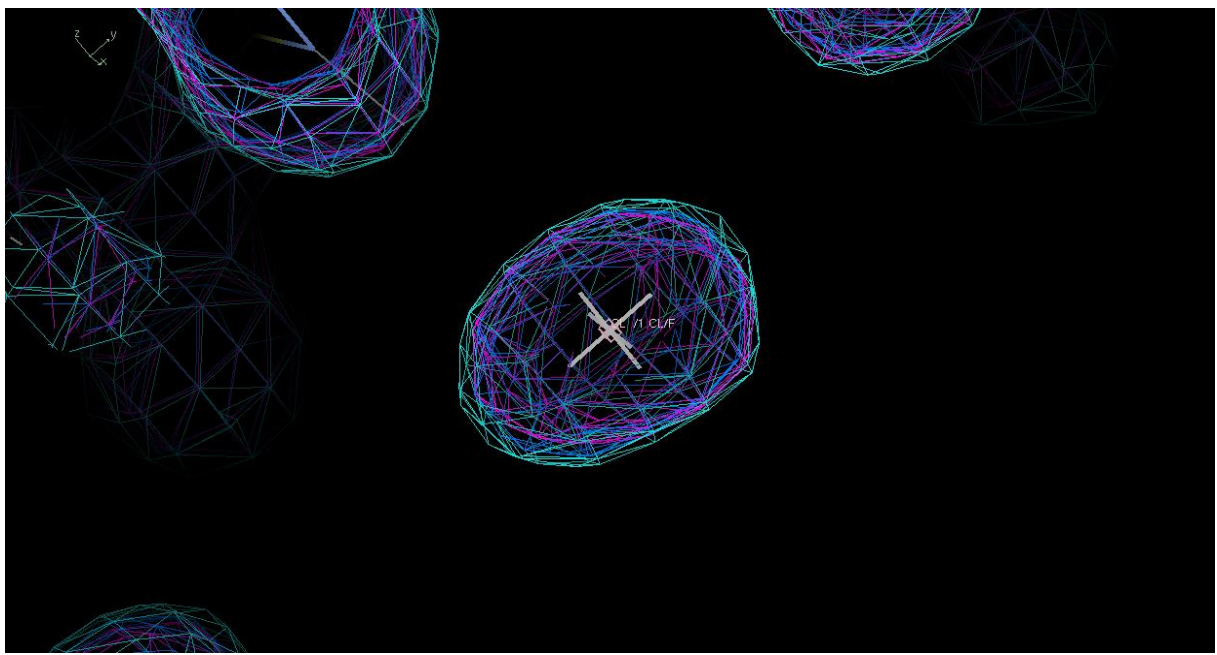

Peak 6:-

Not assigned yet by us. A Br in 4yem ie their Br217.

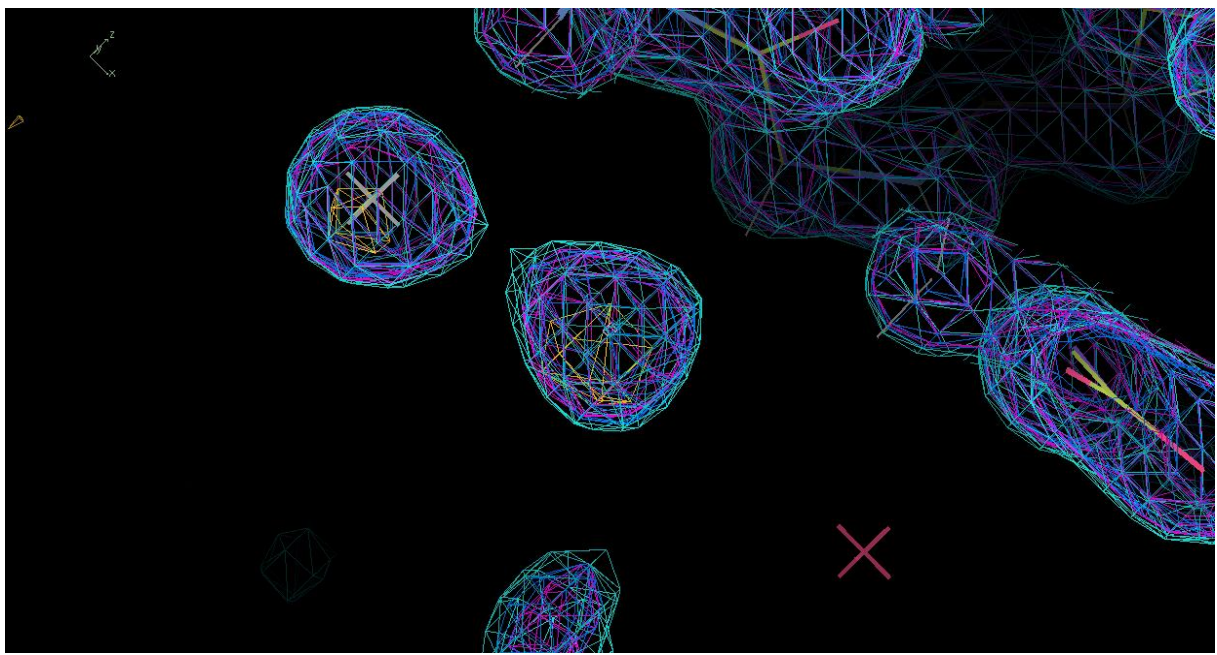

Nb 3.0 Angstrom from another, partially, occupied Br. What is the structural chemistry for this assignment in 4yem?

Peak 7:-

This is the 4yem Br212 analysed by us above ie the cigar shape precludes this assignment.

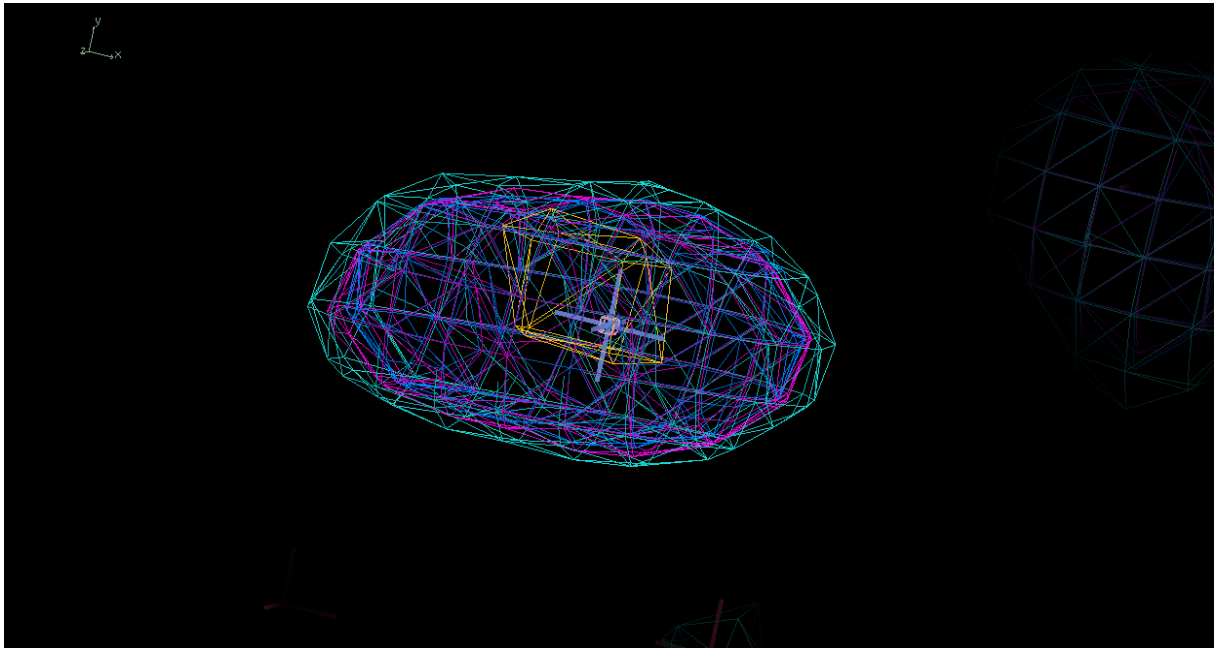

Peak 8:-

This is the Pt 10% occupancy in 4yem. But as remarked above on our revisit the electron density peak shape is elongated and so an assignment has not been made.

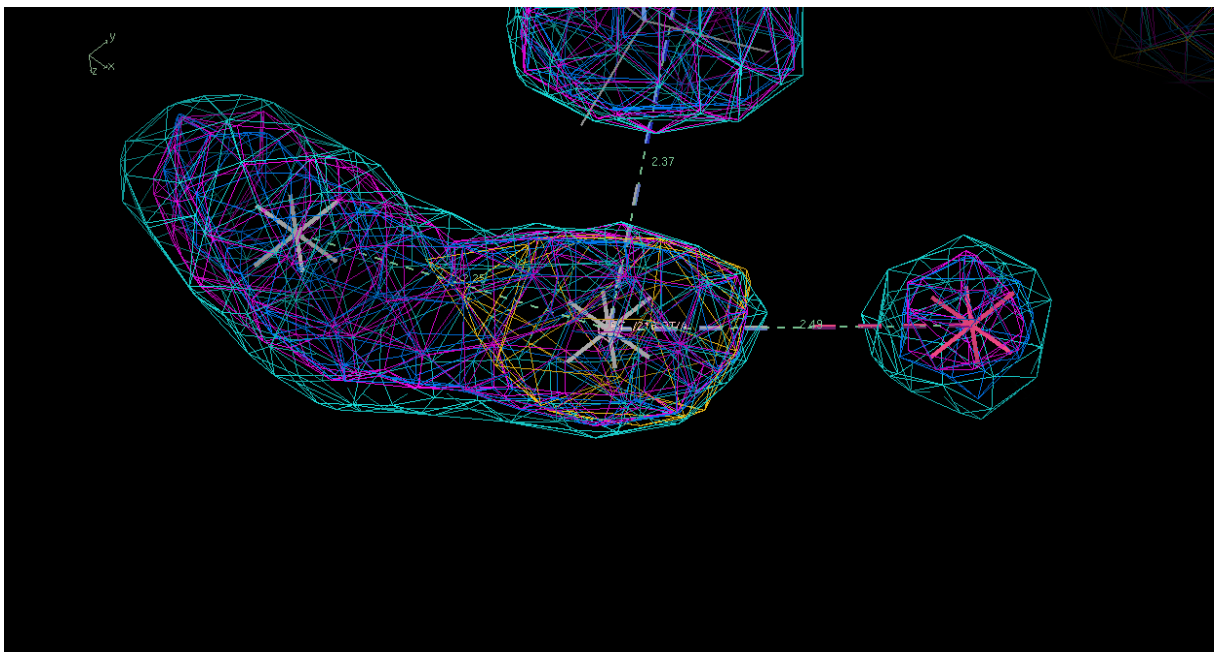

Peak 9:-

This is W395 70% occupancy and a B factor quite low at 17 in 4yem. But the density also has shape; what is it? Unsure so made no assignment.

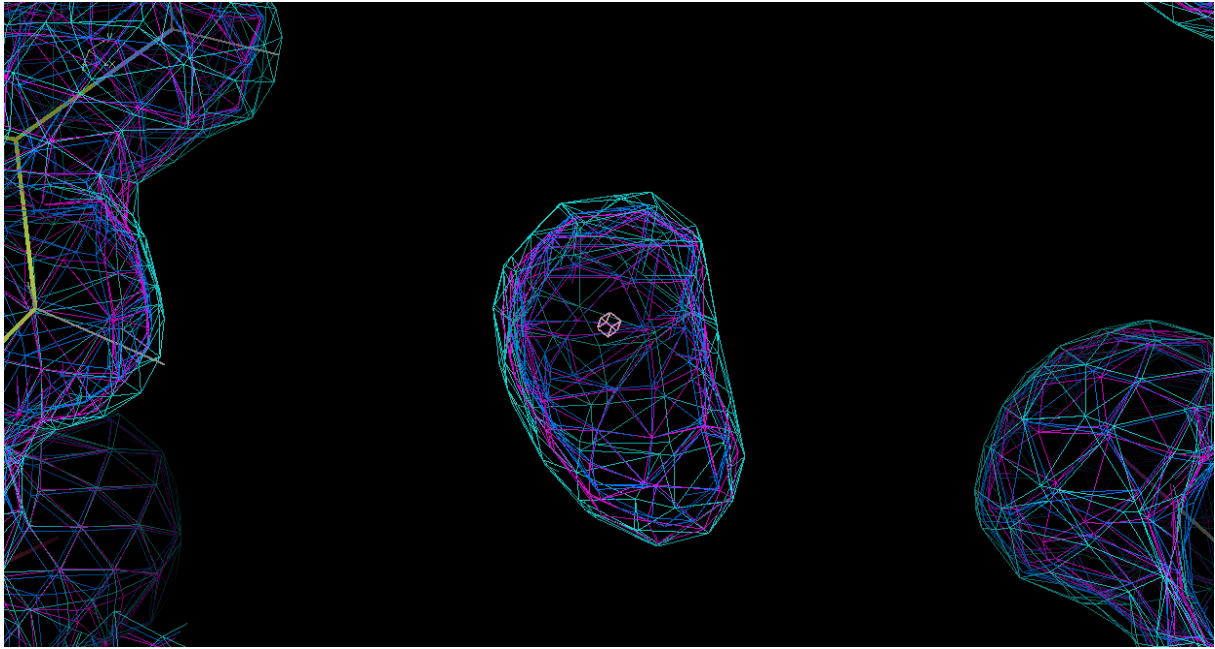

Peak 10:-

This is W392 in 4yem but the electron density has shape. What is it? Unsure so no assignment made.

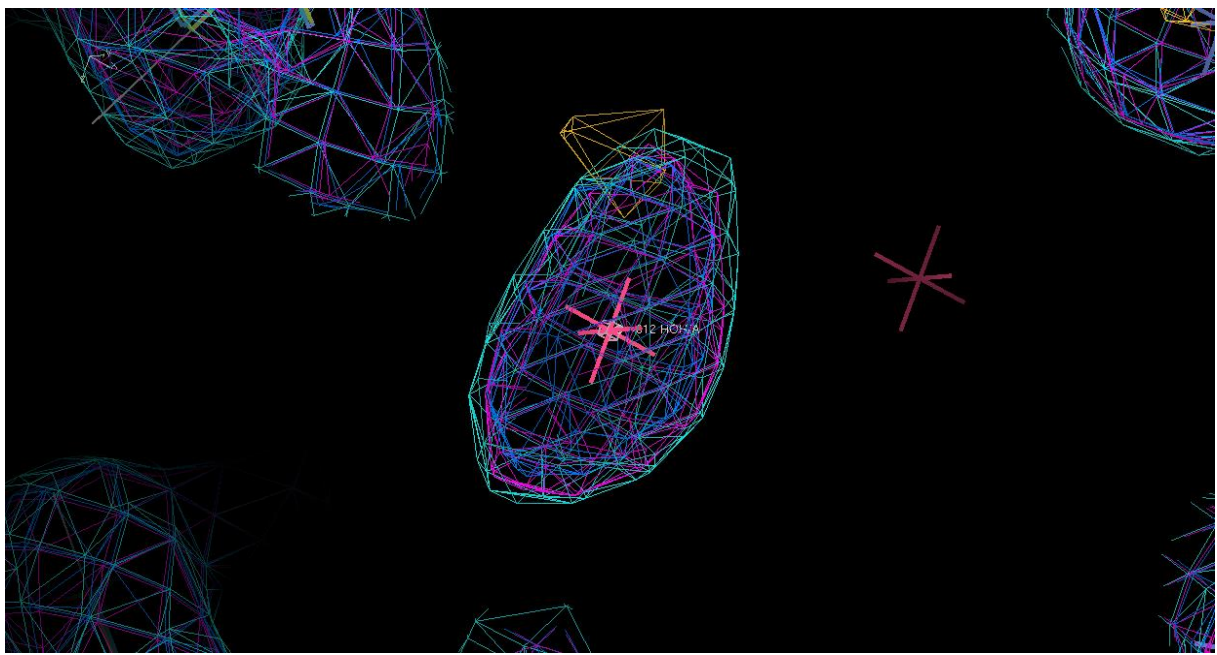

Peak 11:-

This might be a water; it is a bit close to the 30% occupancy Arg68 side chain so could be a 70% bound water. Is that really plausible? Not assigned.  :-

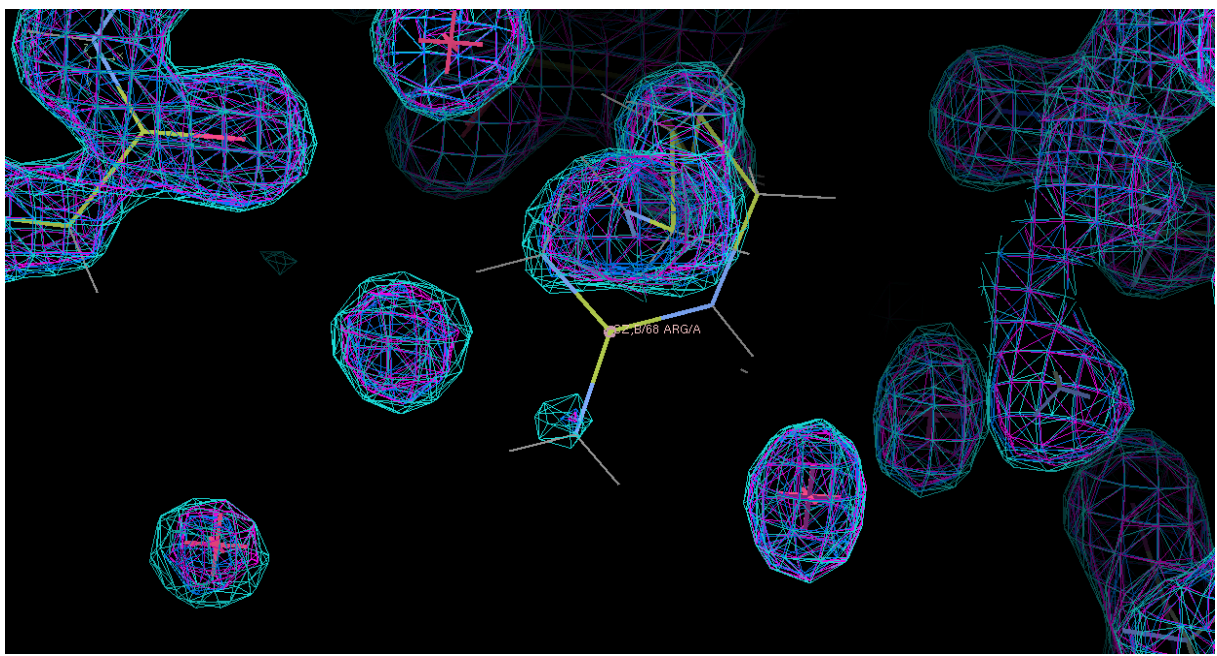

Peak 12:-

Now assigned as a Br (Br H) following the review above:-

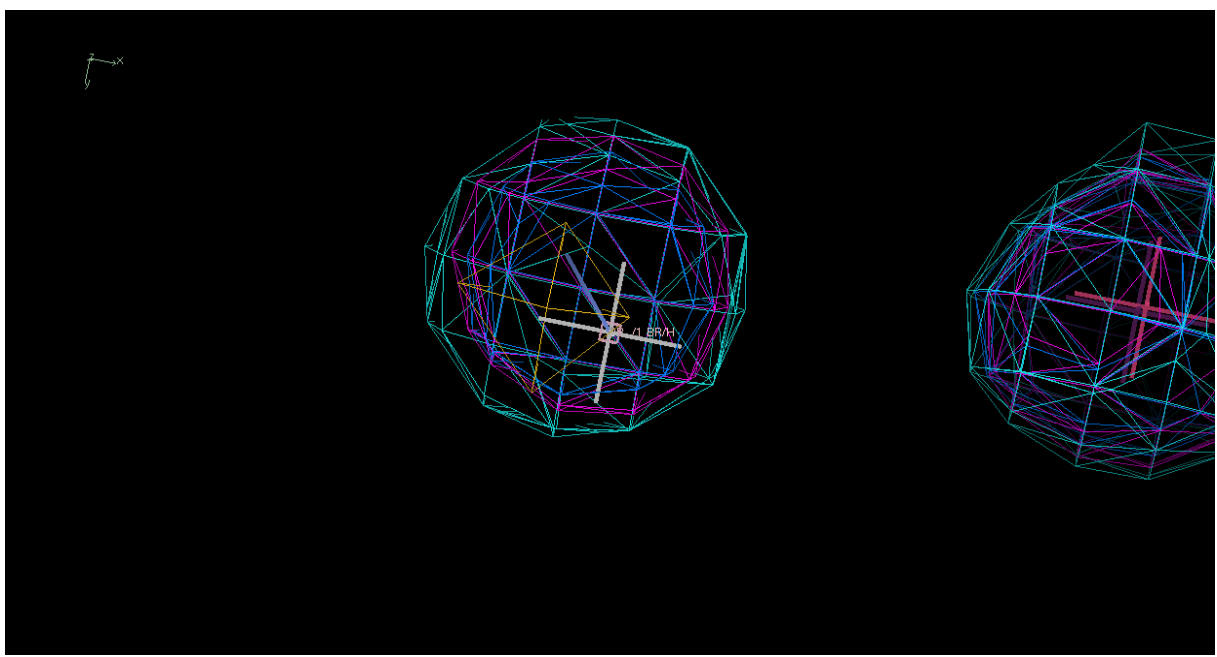

Peak 13:-

This is interpreted as a split occupancy Br in 4yem. I am unsure about this assignment. Not assigned:-

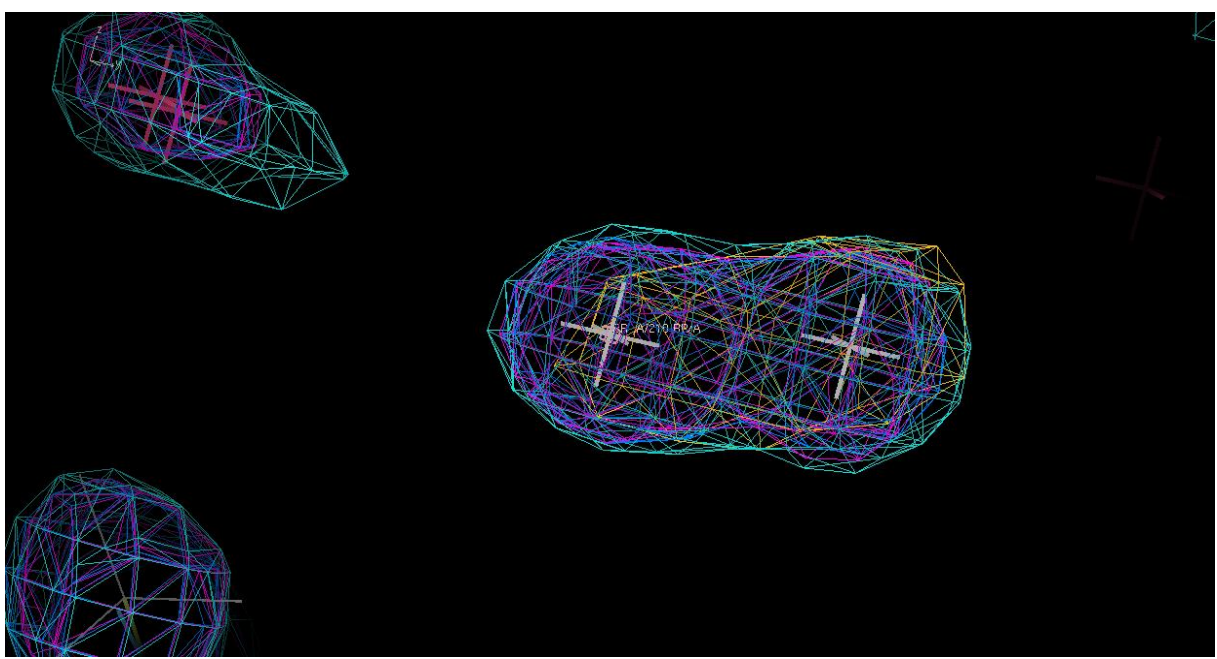

Peak 14:-

This is water 361 in 4yem. Above I decided 'too much shape to be a water'. Not assigned.

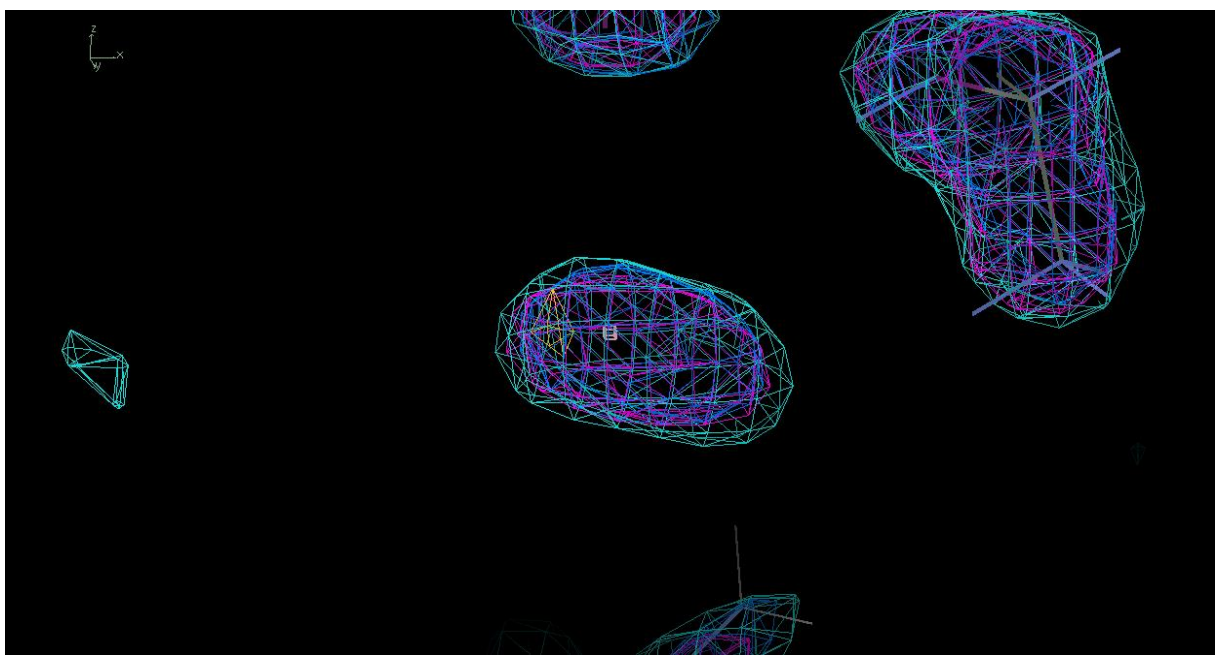

The above top 14 peaks are the ones with consistent agreement across EVAL, XDS and MOSFLM.

**After Cycle 14 the following further checks and actions were taken:-**

Molprobity clashes analysis:-

DMS211 clash and also in negative Fo-Fc (~5 sigma). Delete DMS 211

DMS214 high B factor (72); deleted.

The decisions above on the 4yem bound waters were rechecked since we have made improvements to our model, and this may have led to the possibility of some of the above decisions on the maps to be incorrect. Exceptions were found as follows (ie decisions had to be reversed):-

W378, 403, 469 now do have albeit weak 2Fo-Fc density; added to our model.

W406 to be changed to a Br. [that was stated to have been done above; however I must not have saved it in the latest model properly. Now added.]

**Cycle 14** ie our model W381 is now not in electron density; deleted.

**Cycle 15:-**

A very high B factor for Br L (B 227Å$^2$ !). Deleted.

**Cycle 16:-**

All bound waters, solute and ion assignments look fine ie with map evidence. Output mmcif files for PDB upload (done at January 16<sup>th</sup> 2016; new PDB code 5HMJ (replacing 4xan)).

**Part E Unexplained 'blobs of electron density'**

The Fo-Fc difference map peaks that finally remain are these:-

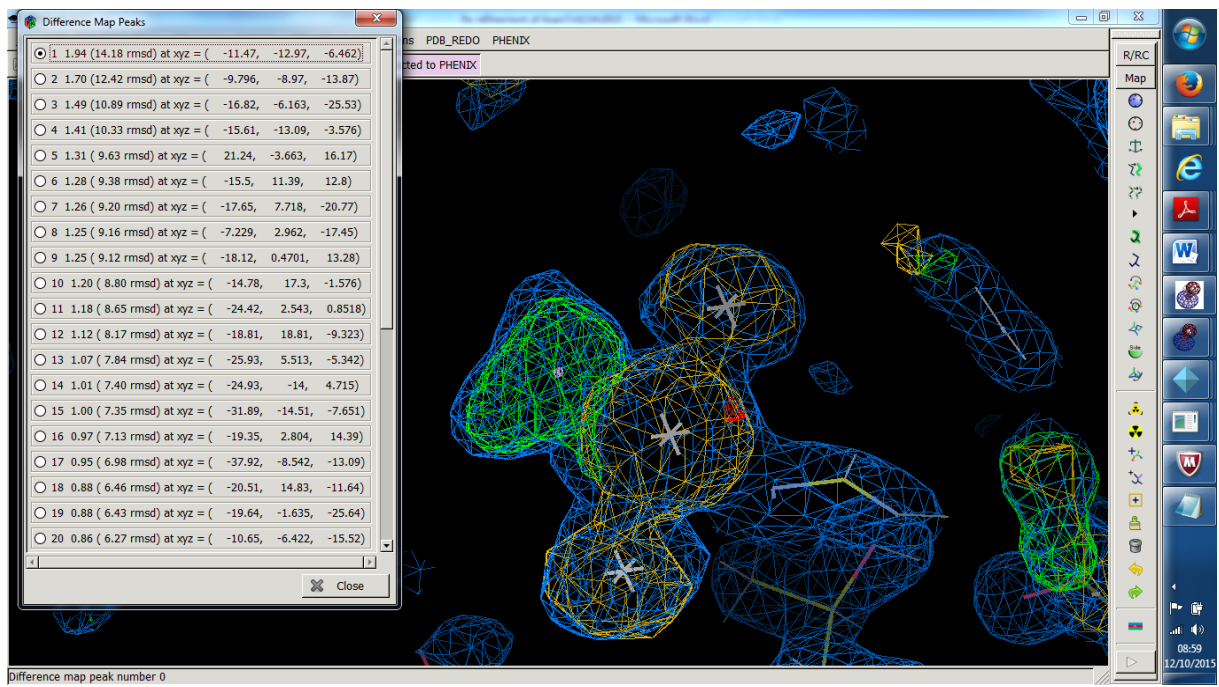

The maps for the top twelve peaks are now shown one by one below. They are very similar to those checked at Cycle 8 above with the exception that Peak 2 in that list is now assigned (as a partial occupancy Br anion).

Peak 1:-

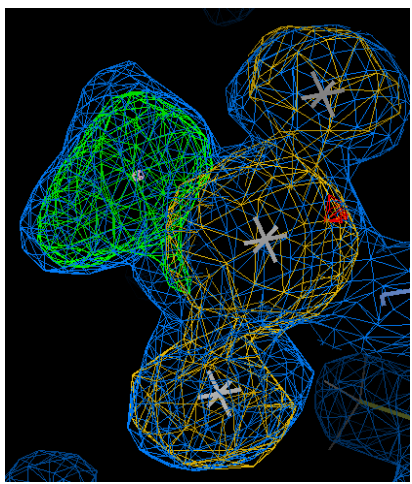

Peak 2:-

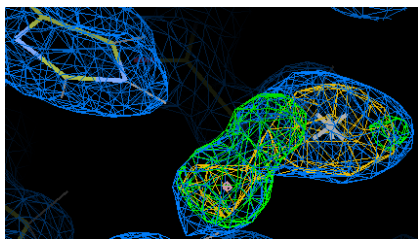

Peak 3:-

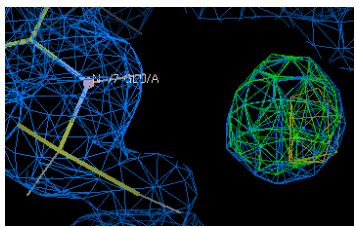

Peak 4:-

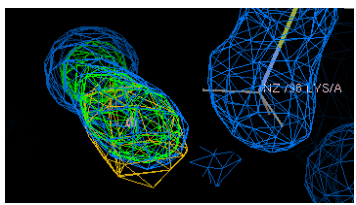

..............................................................

Peak 5:-

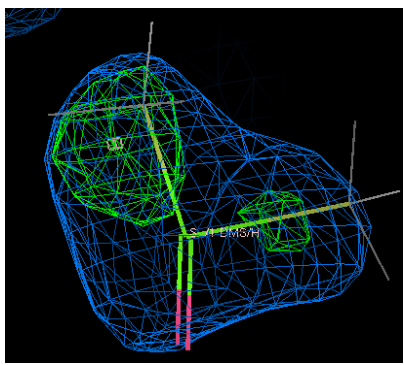

..............................................................

Peak 6:-

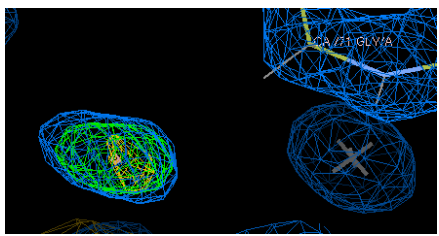

Peak 7:-

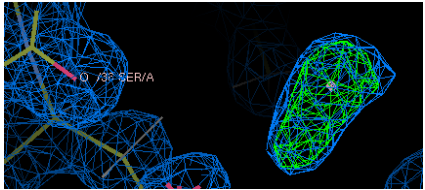

Peak 8:-

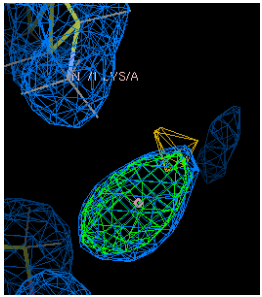

Peak 9:-

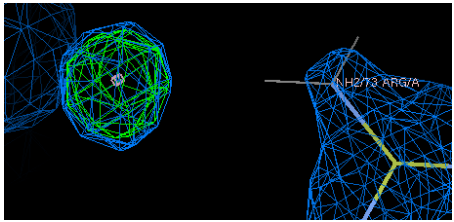

Peak 10:-

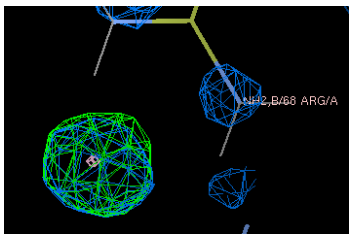

Peak 11:-

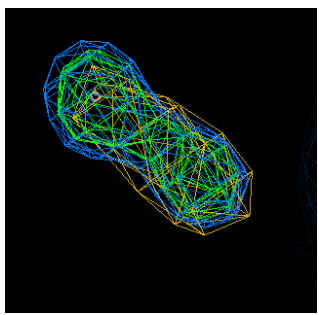

Peak 12:-

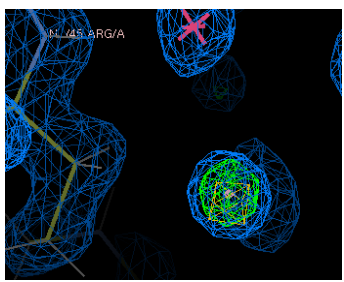

## Supplementary 3

The PDB's Validation report section 6.4 on our 'Cycle 16' model highlighted concern with respect to 'DMS (308)' (formerly DMS211). The 'DMS (308)' solute molecule was highlighted by the PDB as having a high, ie poor, LLDF of 11.26. Figure S3.1 below shows the electron density evidence for it. We also comment that DMS 214 was similar. 4yem has also got both these as DMSO. We retain them.

Figure S3.1 The 2Fo-Fc (blue) and Fo-Fc (green) electron density for the 'DMS (308)' molecule.

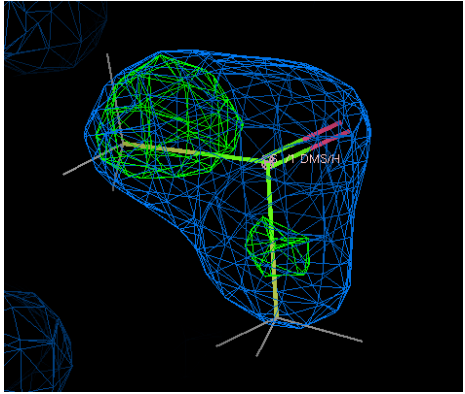